\documentclass{emulateapj}
\usepackage{graphicx}
\usepackage{color}

\newcommand{\km}{\,\mbox{km}\,\mbox{s}^{-1}}
\newcommand{\cm}{\,\mbox{cm}^{-3}}
\def\Ha{H$\alpha$}
\def\Hb{H$\beta$}

\shorttitle{Stars and ionized gas  in NGC~7743}
\shortauthors{Katkov et al.}

\begin{document}
\slugcomment{{\it The Astronomical Journal, accepted}}

\title{Stars and ionized gas in S0 galaxy NGC~7743: an  inclined large-scale
gaseous disk.\footnote{Based on the observations collected with the 6-m  telescope
of the Special Astrophysical Observatory  of the   Russian Academy of Sciences
which is operated under the financial support of the Science Department of Russia
(registration number 01-43)}}

\author{Ivan Yu. Katkov}
\affil{Sternberg Astronomical Institute, Moscow, 119992 Russia}
\email{katkov.ivan@gmail.com}

\author{Alexei V. Moiseev}
\affil{Special Astrophysical Observatory, Russian Academy of Sciences, Nizhnii
Arkhyz,  Karachaevo-Cherkesskaya Republic, 369167 Russia}
\email{moisav@gmail.com}

\author{Olga K. Sil'chenko}
\affil{Sternberg Astronomical Institute, Moscow, 119992 Russia\\
and Isaac Newton Institute of Chile, Moscow Branch}
\email{olga@sai.msu.su}

\begin{abstract}
We  used deep  long-slit spectra and integral-field spectral data to study the stars
and ionized gas kinematics and stellar population properties in the lenticular
barred galaxy NGC~7743. We have shown that ionized gas at the distances larger
than 1.5 kpc from the nucleus settles in the disk which is significantly
inclined to the stellar disk  of the galaxy.
 Making different assumptions about the geometry of the disks and involving different sets of emission lines
into the fitting, under the assumption of thin flat disk circular rotation, we obtain the full possible range of
angle between the disks to be $34\pm9\degr$ or $77\pm9\degr$.
The most probable origin of the inclined  disk is the external
gas accretion from a satellite, orbiting the
host galaxy with a corresponding angular momentum direction. The published data
on the HI distribution around NGC~7743 suggest that the galaxy has a gas-rich
environment. The emission-line ratio diagrams imply the domination of shock waves in
the ionization state of the gaseous disk, whereas the contribution of
photoionization by recent star formation seems to be negligible.   In some parts of the disk a difference between  the velocities of the gas emitting in the forbidden lines and in the Balmer lines is detected. It may be caused by the fact that the inclined disk is mainly shock-excited, whereas some fraction of the Balmer-line emission is produced by a small amount of gas excited by young stars in the main stellar disk of NGC 7743.
In the circumnuclear region ($R< 200$ pc) some evidences of the AGN jet interaction
with an ambient interstellar medium were found.
\end{abstract}


\section{Introduction}

Interactions play an important role in the galaxy evolution. Even a minor merger
with the mass ratio less  than $1/5$--$1/10$ which does not disturb the overall
structure of the galaxy disk, may cause a gas concentration in its central region,
triggering the nuclear activity or a nuclear starburst. According to the
hierarchical paradigm, such events happened many times during a galaxy lifetime,
however, it is difficult to detect minor merging footprints, even the recent
ones, against the background of a high surface brightness galaxy. Deep images,
revealing low-contrast tidal features  \citep*[see, for
example][]{Martinez-Delgado2010, Smirnova2010b}, or  detailed studies of
stellar populations and  internal kinematics of the galaxy are required.

The consequences of a minor merger may be quite diverse. Own gas of a
larger galaxy may be disturbed by a satellite intrusion, hence it may inflow into
the center, resulting in a strong gas compression in the nucleus and a subsequent
nuclear star formation burst. A rather young current age
of the nuclear star population inside an older bulge may be the signature of
such event. If the initial direction of the orbital angular momentum of the
satellite significantly differed from the main galaxy disk rotation momentum,
then  we may now have a situation when some fraction of  stars or gas clouds
in the galaxy are rotating on orbits tilted (or orthogonal) to the main galactic
disk, or even counter-rotating within it.   Long-slit and integral-field
spectral observations gave a number of evidences for the kinematical misalignments
in circumnuclear regions of early-type disk galaxies including the cases of
kinematically decoupled cores. Numerous examples and detailed discussions and
references can be found, for instance, in the papers by
\citet{weall89,Corsini2003, Moiseev2004,Sarzi2006,Coccato2007}.  Unfortunately,
observational evidences  of large-scale (beyond one kiloparsec central region)
kinematically decoupled subsystems are still quite rare \citep[see][and
references therein]{Silchenko2009}. Therefore, every new example of the
similarly peculiar objects is interesting.

In the present paper, we demonstrate  that  the main fraction of the ionized gas
in NGC~7743 rotates on the orbits, considerably inclined  to the main stellar
disk of the galaxy, what may be a result of   external gas accretion or tidal
destruction of a gas-rich small companion.
The paper is organized as follows:  Section~\ref{sec_n7743} gives an overview of
the literature on the previous studies of the  galaxy; Section~\ref{sec_obs}
describes the spectroscopic observations and data reduction process; in
Section~\ref{sec_stars}  the properties of the stellar population (kinematics,
age, and metallicity) are considered;  Section~\ref{sec_gas} contains the study
of gas kinematics and ionization state;    Section~\ref{sec_dis}  includes an
overall discussion of the structure and kinematics of NGC~7743.

\begin{figure}
\includegraphics[width=0.5\textwidth]{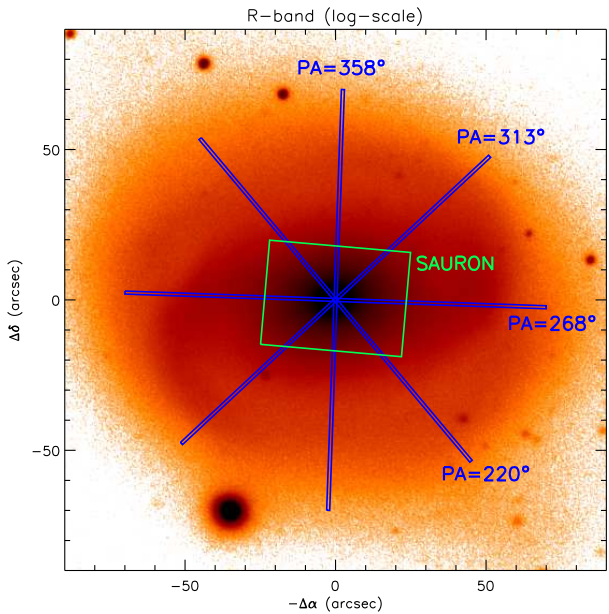}
    \caption{The positions of the SCORPIO slits and the SAURON mosaic field of
view are overlapped onto the $R$-band image of the galaxy NGC~7743 from
\citet{Moiseev2004}.}
    \label{fig_r_slits_sau}
\end{figure}

\section{NGC~7743: What was known  before?}
\label{sec_n7743}

NGC~7743 is a barred early-type galaxy (NED morphological type  (R)SB0$^+$(s))
with a total blue luminosity of $M_B=-19.4$ (according to  the HyperLeda
database\footnote{\textrm{http://leda.univ-lyon1.fr}}). Following
\citet{Jensen2003}, we adopt the distance to the galaxy to be 19.2~Mpc, that
corresponds to a linear scale of $93\,\mbox{pc}\,\mbox{arcsec}^{-1}$.

A smooth two-armed spiral structure without any traces of star formation
dominates in that optical images of the galaxy.   According to \citet*{Ho1997} the
galaxy has an active nucleus of the Sy2 type. Radio observations reveal  a
compact (under the  beam$=1\arcsec -5$\arcsec) non-thermal source in the nucleus
\citep{Nagar1993,HoUlvestad2001}.  However, the nuclear activity  is not very
high. \citet{Alonso-Herrero2000}  using their spectral observations classified
NGC~7743 as a `low-luminosity AGN' and noted that the optical emission line
ratios  correspond to the boundary case between Sy2 and LINER activity types.
The X-ray observations by \citet{Terashima2002} also suggest a relatively low
activity of the NGC~7743 nucleus, compared with other Seyfert galaxies from
their sample.

The HST imaging reveals a complex  structure of the circumnuclear region
($R<300$--$500$ pc) where emission knots and several curved dust lanes are
observed  \citep{ReganMulchaey1999}. \citet{Martini2003} described this
structure as a loosely wound circumnuclear spiral. \citet*{Moiseev2004}  used
integral-field spectroscopy to study the inner region morphology and kinematics.
They suggested that a turn of the innermost isophotes relates with the
circumnuclear dust spiral, rather than with a triaxial bulge as it was  claimed earlier.
Despite the relatively low  intensity  of  emission lines, they succeeded to
obtain some conclusions about the ionized gas kinematics. In the region of
$R<2\arcsec -4\arcsec$  (190--370 pc) the gas motions agree in general with
the stellar disk rotation, but non-circular ionized gas motions were also
detected  locally to the south from the nucleus. \citet{Moiseev2004} have as well
supposed that the inner part of the galaxy disk can be tilted with respect to
the outer disk.

Radio observations by \citet{DuprieSchneider1996}  demonstrated  a very low
content of  neutral gas (near their detection limit) for the NGC~7743 disk
through the  $\sim 3\arcmin$ beam, whereas two separate HI clouds were
discovered in the tight neighborhood of the galaxy having the masses of
$M_{HI}=6.4\times10^7\,M_\odot$ and $4.5\times10^8\,M_\odot$. Their systemic
velocities are $1610\km$ and $1509\km$, respectively, that is close to the
NGC~7743 systemic velocity which is $1710\km$ (HyperLeda). According to the NED
database these clouds are associated  with the galaxies  KUG~2341+097  and
LSBC~F750-04, with projected  distances of 8\farcm3 (46 kpc) and 11\farcm0 (61 kpc)
from NGC~7743.  The absolute magnitude of the brighter galaxy (KUG~2341+097 ) is
 about $M_B=-16$.   Therefore NGC~7743 is surrounded  by the satellites which
altogether contain the neutral hydrogen amount  by one order larger compared to the
disk  of the main galaxy. At the same time   \citet{Maiolino1997} from their
radio observations with a  beam$=55\arcsec$ detected  the CO emission,
corresponding to the total mass of the molecular gas in the galaxy disk of about
$1.3 \times10^8\,M_\odot$, which is comparable with the HI mass in the local environment of NGC~7743.

\section{Spectral observations and data reduction}
\label{sec_obs}

 In our study of NGC~7743, we have used two types of spectral data.
Long-slit spectroscopy with exposures long enough allows to reach very
outer parts of the galaxy disk; however, one-dimensional character of the
long-slit data restricts consideration of any azimuthal variations.
Integral-field spectroscopy (so called `3D spectroscopy') provides
two-dimensional mapping of the kinematical and stellar population characteristics
but only for the central part of the galaxy due to the limited field-of-view.

\begin{table*}
\footnotesize
\caption{Log of the observations}\label{tab1}
\begin{tabular}{lcccccc}
\tableline
 Instrument & Date        &  $PA$ (deg)          & Sp. range (\AA)  &Sp.
resolution (\AA)	& Exp. time (s)   &  Seeing (\arcsec)\\
\tableline
    SCORPIO & 2007~Sep~21 &  268                  &      4800--5550  &      2.5
    	   	&    $9\times1200$ 		&   1.9\\
            & 2009~Jan~22 &  268                  &      6100--7100  &      3.1
       		&    $2\times500$  		&   2.4\\
            & 2007~Sep~21 &  358                  &      4800--5550  &      2.5
       		&    $9\times1200$ 		&   1.8\\
            & 2009~Jan~22 &  358                  &      6100--7100  &      3.1
       		&    $2\times400$  		&   3.0\\
            & 2008~Dec~19 &  313                  &      4800--5550  &      2.5
       		&    $8\times1200$ 		&   1.2\\
            & 2009~Oct~15 &  220                  &      6100--7100  &      3.1
       		&    $10\times1200$		&   1.6\\
\tableline
     SAURON & 2007~Aug~14 &  355 (Pos. 1)     &      4800--5400  &      4.8
   		&    $2\times1800$ 		&   2\\	
            & 2007~Aug~14 &  355 (Pos. 2)     &      4800--5400  &      4.8
   		&    $2\times1800$ 		&   2\\
\tableline
\end{tabular}
\end{table*}

\subsection{SCORPIO long-slit observations.}

The   spectral observations were made at the prime focus of the SAO RAS 6-m
telescope with the multimode
focal reducer SCORPIO \citep{AfanasievMoiseev2005}. The slit had  6\farcm1  in
length and 1\arcsec\ in width.  The  $2048\times2048$ EEV 42-40 CCD provided the
scale of $0.35\,\mbox{arcsec}\,\mbox{px}^{-1}$ along the slit. The log of
observations is given in Table~\ref{tab1}, the slits positions onto the galaxy
image are shown in Fig.~\ref{fig_r_slits_sau}.

The starting observations in 2007 were performed in order to study the stellar
kinematics (P.I.  of the proposal was Anatoly Zasov), therefore, the `green'
spectral range of 4800--5550\AA\,   was chosen. It contains strong stellar
absorption line features as well as the ionized gas  emission lines -- \Hb\,
and  $\mbox{[OIII]}\lambda\lambda4959,5007$. After the preliminary analysis of these
data a difference in the kinematics of gas and stars was found, and  we carried
out additional observations in the  `red' spectral domain of 6100--7100\AA\,
including  the brighter emission lines --  \Ha,
$\mbox{[N\,II]}\lambda\lambda6548,6583$, and
$\mbox{[S\,II]}\lambda\lambda6716,6731$.

\begin{figure*}
\centerline{
 \parbox[r]{0.25\textwidth }{~~~~~~~$PA=220\degr$\\
\includegraphics[width=0.25\textwidth]{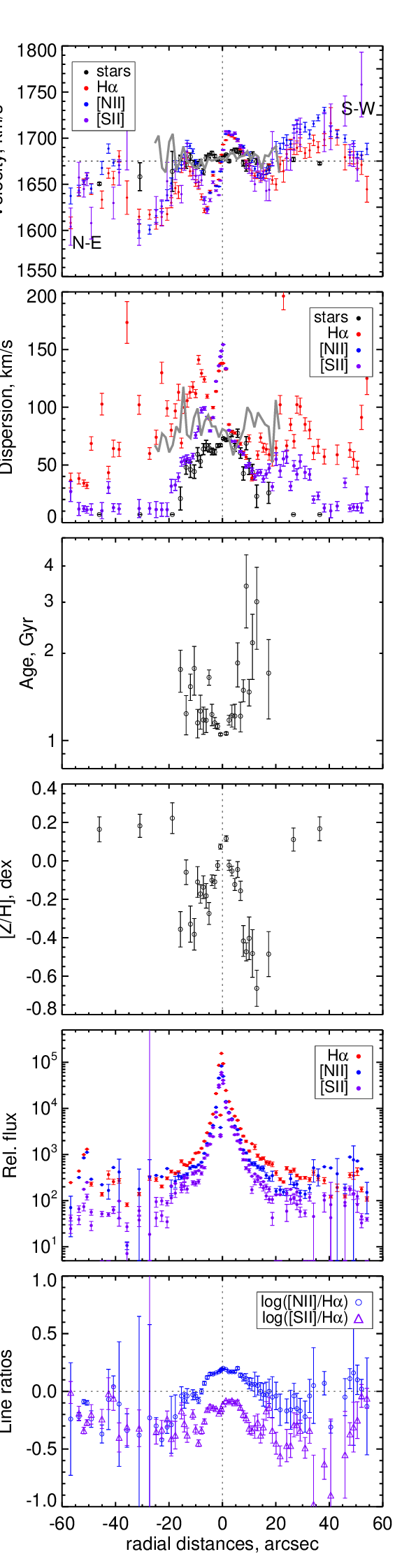}}
 \parbox[r]{0.25\textwidth }{~~~~~~~$PA=268\degr$\\
\includegraphics[width=0.25\textwidth]{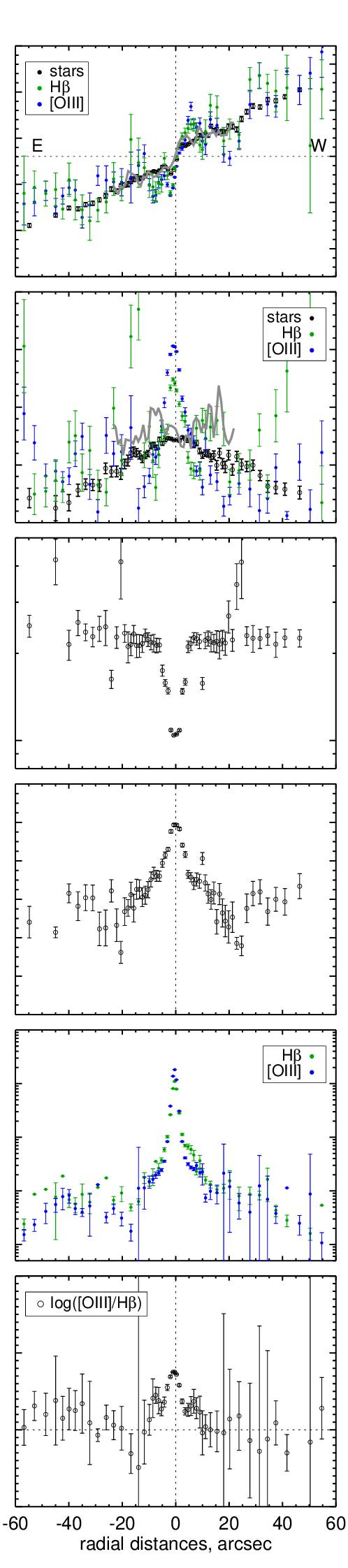}}
 \parbox[r]{0.25\textwidth }{~~~~~~~$PA=313\degr$\\
\includegraphics[width=0.25\textwidth]{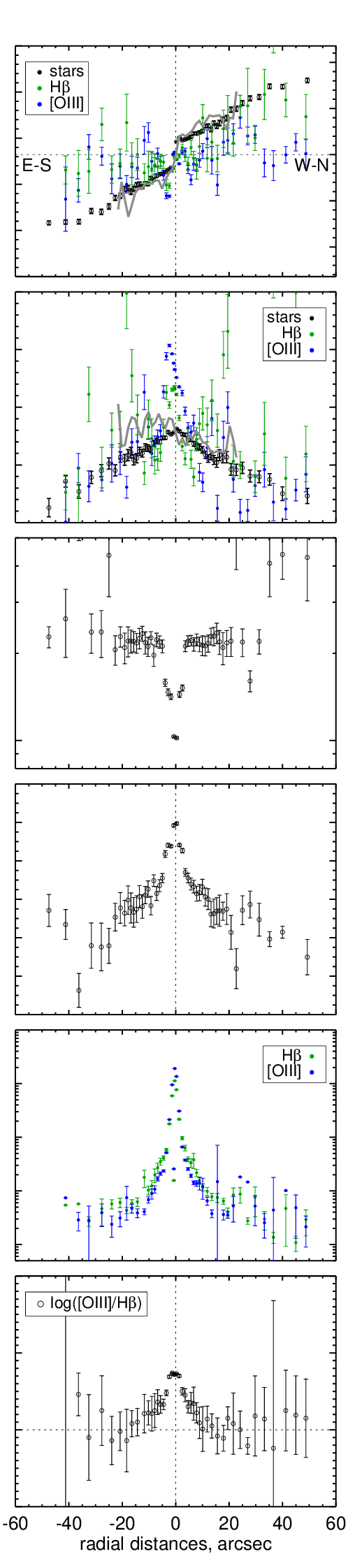}}
 \parbox[r]{0.25\textwidth }{~~~~~~~$PA=358\degr$\\
\includegraphics[width=0.25\textwidth]{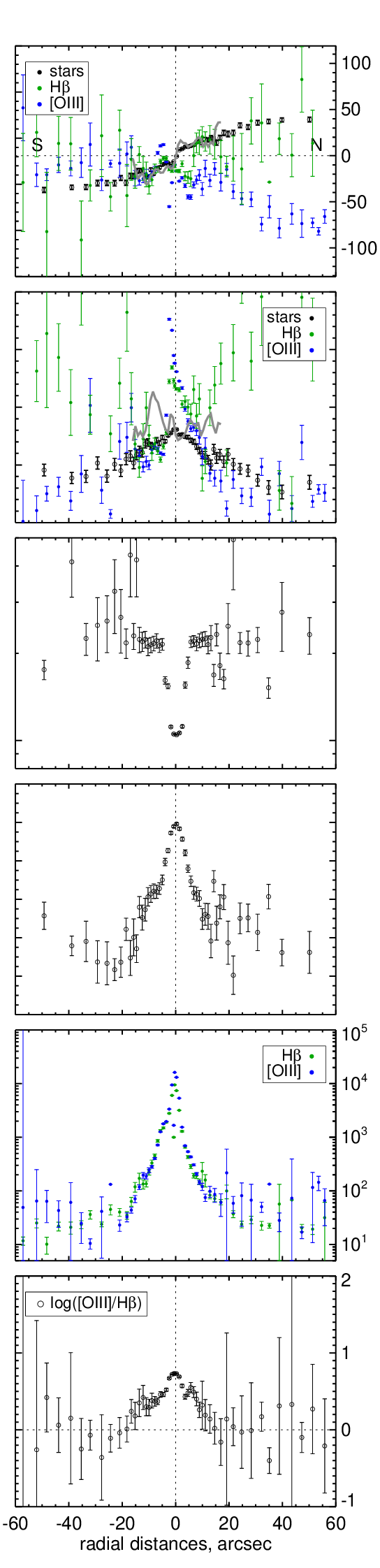}}
}
  \caption{The results of the long-slit observations. Each column corresponds
to the slit position angle labelled at the top.    Data for   $PA=220\degr$  are
shown for the  `red'  spectral range, in other cases we present the results only
for the `green' spectral domain. From   top to bottom we show  the radial
distributions of the line-of-sight velocity, the velocity dispersion, the age
and metallicity of the stellar component, the intensity of the emission lines
(in arbitrary units) and the line intensity ratios.  Grey lines correspond
to the pseudo slits extracted from the IFU data.}
  \label{fig_longslit_results}
\end{figure*}

The  spectra were reduced  by the standard way using our IDL-based software
\citep[see, for instance,][]{Zasov2008}. For the night sky spectrum subtraction,  we
applied  an advanced technique  taking into account the variations of the Line
Spread Function (LSF) along the slit \citep{Katkov_inprep}. The model of the LSF
was derived from the twilight sky spectra observed the same nights as the galaxy.
These spectra were binned by 10 pixels along the slit, the wavelength  range was
divided into six segments. The LSF parameters were estimated over every segment
by fitting the observed spectrum by a broadened high-resolution spectrum of the
Sun using the  ULySS software\footnote{http://ulyss.univ-lyon1.fr/}
\citep{Koleva2008ssp, Koleva2009ulyss} adapted for the SCORPIO data.

The further data analysis  included following steps:
\begin{itemize}
 \item Adaptive  binning along the slit -- we summed spectra in order to
achieve   the minimal value of signal-to-noise ratio  $S/N=20$--$30$.
 \item  Masking the emission line regions and fitting the binned galaxy
spectra with a simple stellar population (SSP)
 models PEGASE.HR \citep{LaBorgne2004} convolved with a parametric line-of-sight
velocity distribution (LOSVD) using ULySS software.
The resulting stellar physical parameters are the line-of-sight velocity -- $v$,
velocity dispersion  -- $\sigma$, higher-order Gauss-Hermite moments $h_3$, $h_4$; and the SSP-equivalent
luminosity-weighted parameters: the age $T$ and
metallicity [Z/H]. The stellar population models were constructed with the local
star spectra library, therefore, the element abundance ratios, [Mg/Fe] in particular,  cannot be free variables. While we deal with a rather high model metallicity,
[Fe/H]$>-0.4$, the model alpha-element-to-iron ratio is necessarily close to the
solar one due to the fixed chemical properties of the stars in the Sun' neighborhood. For a more detailed description see \citet{Koleva2008ssp, Koleva2009ulyss}.
\item The subtraction of the stellar contribution from non-binned
spectra using the model stellar spectra evaluated in every bin at the previous
step, and normalized here by a polynomial approximation of the surface brightness distribution. The result
is a non-binned ionized gas spectrum.
 \item Adaptive binning of the emission lines spectra. The bins were formed
under the condition that $S/N>10$ for the targeted emission line.
\item The emission lines fitting using the Gaussians convolved with the LSF. For the
doublets of $\mbox{[OIII]}$, $\mbox{[NII]}$, and $\mbox{[SII]}$ the restframe separation
between components, their widths and intensity ratios (except the sulfur
lines) were fixed.
\end{itemize}

It should be noted that   the cross-sections $PA=268\degr$ and  $358\degr$ in
the `red' spectral range were observed with relatively low exposure times, and
emission lines measurements in these spectra gave no any additional
information about the ionized gas kinematics with respect to the deep spectra in
the `green' spectral domain. Therefore we have used the measurements of the
emission lines  \Ha,  $\mbox{[NII]}$ and $\mbox{[SII]}$ in these $PAs$ only to
derive the ionization state diagrams. On the contrary, in  $PA=220\degr$ (the galaxy
minor axis) we took very deep `red' spectra, where the ionized gas emission lines
were detected at large distances from the nucleus. At the same time the
estimations of the stellar population parameters in the `red' domain are
uncertain, opposite to the `green' spectra.

The kinematical profiles for the stellar and gaseous components, the emission
lines ratios  and radial profiles of the stellar population parameters are shown
in Fig.~\ref{fig_longslit_results}.

\subsection{SAURON integral-field spectroscopy.}

We also  involved the data from the 4.2-m William Herschel Telescope
integral-field spectrograph SAURON \citep{Bacon2001}  obtained in the
framework of the ATLAS3D survey \citep{Cappellari2011}. The raw data have
been retrieved by us from the open Isaac Newton Group Archive, which is
maintained as a part of the CASU Astronomical Data Centre at the Institute of
Astronomy, Cambridge. NGC~7743 was observed in two positions of the SAURON
lenslet array $33\arcsec \times 41\arcsec$ in size centered onto the opposite
sides relative to the galactic nucleus. The resulting central part of the galaxy
observed was $35\arcsec \times 50\arcsec$ (see  Fig.~\ref{fig_r_slits_sau}) with
 0\farcs94 sampling. For our analysis we have used the scientific-ready spectral
data cubes which have been presented earlier by
\citet{SilchenkoChilingarian2011}.

\begin{figure*}
     \centerline{
       \includegraphics[width=0.33\textwidth]{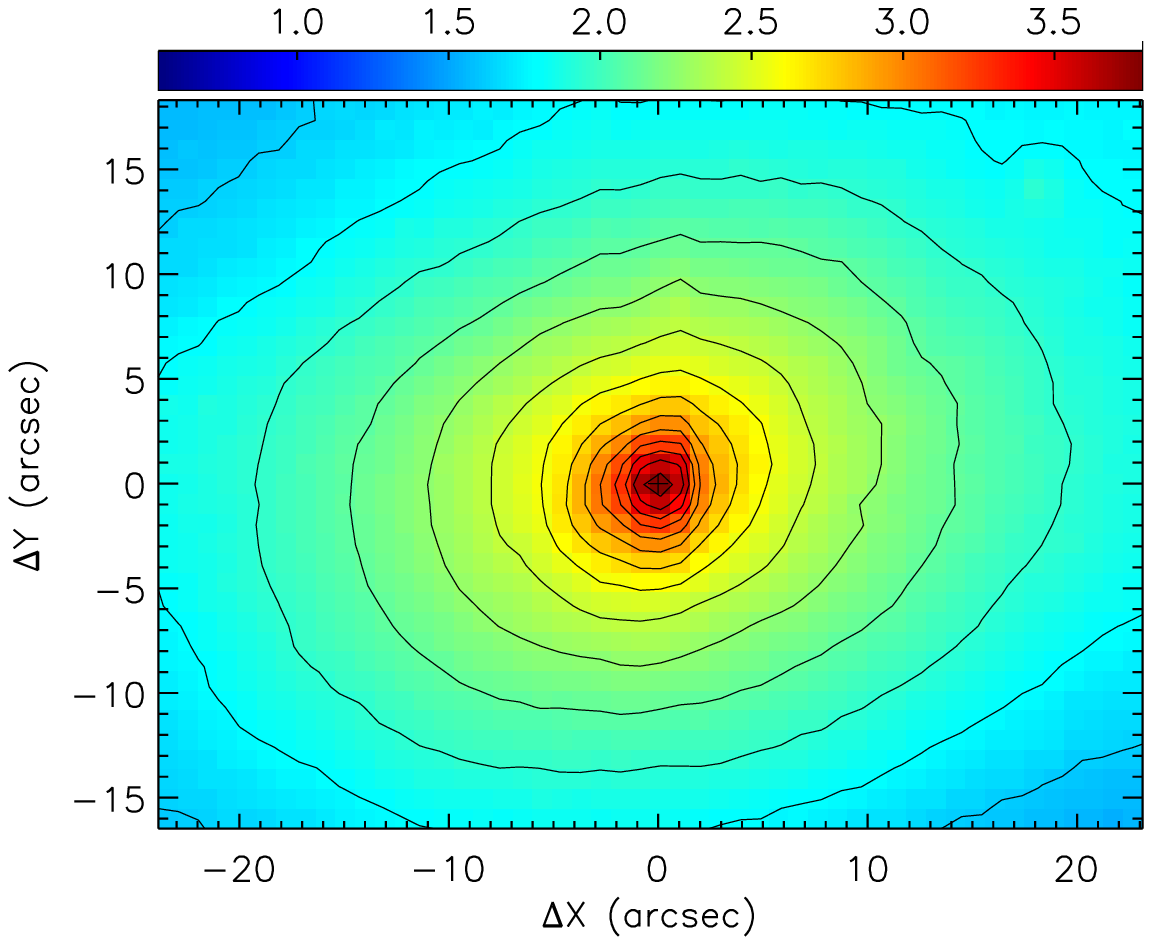}
       \includegraphics[width=0.33\textwidth]{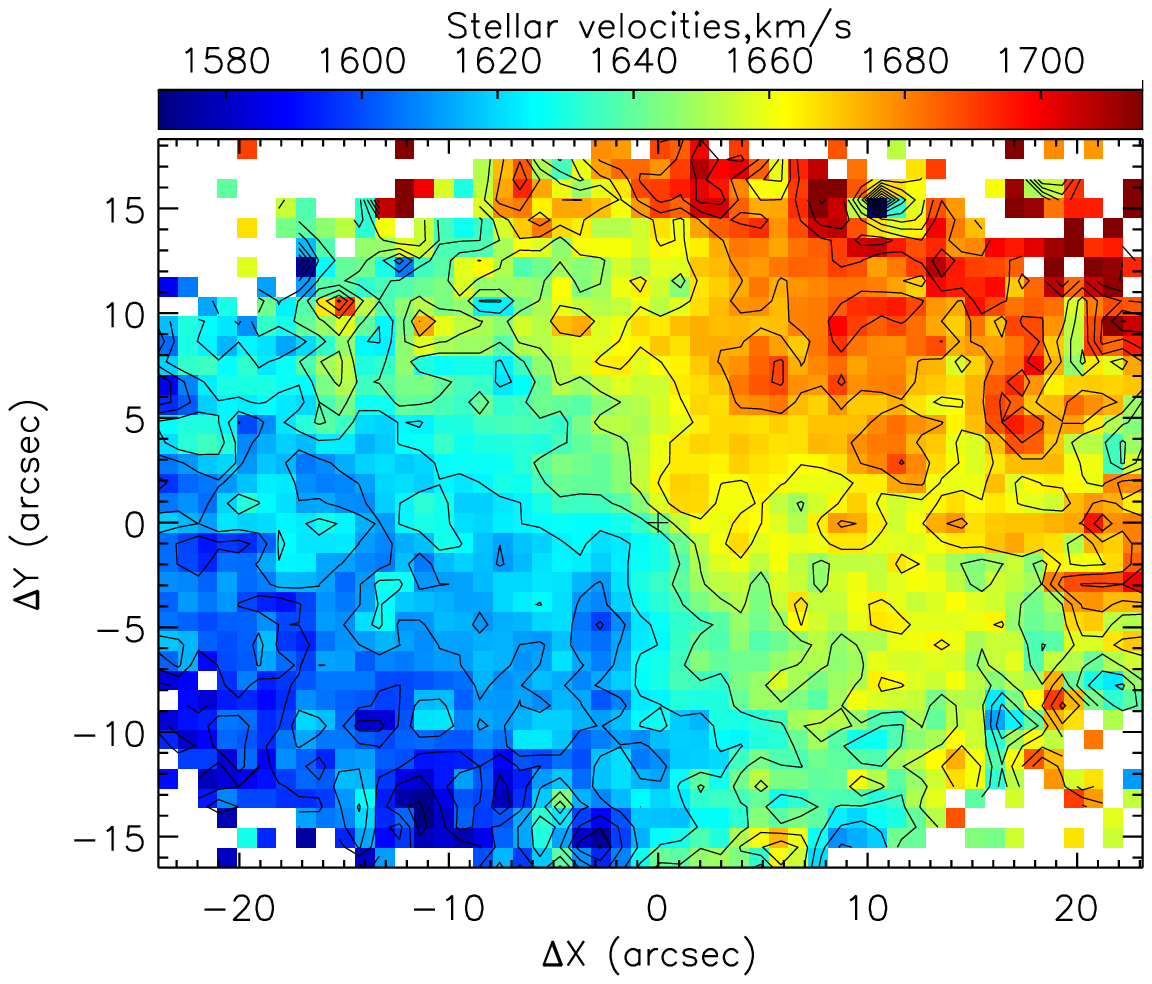}
       \includegraphics[width=0.33\textwidth]{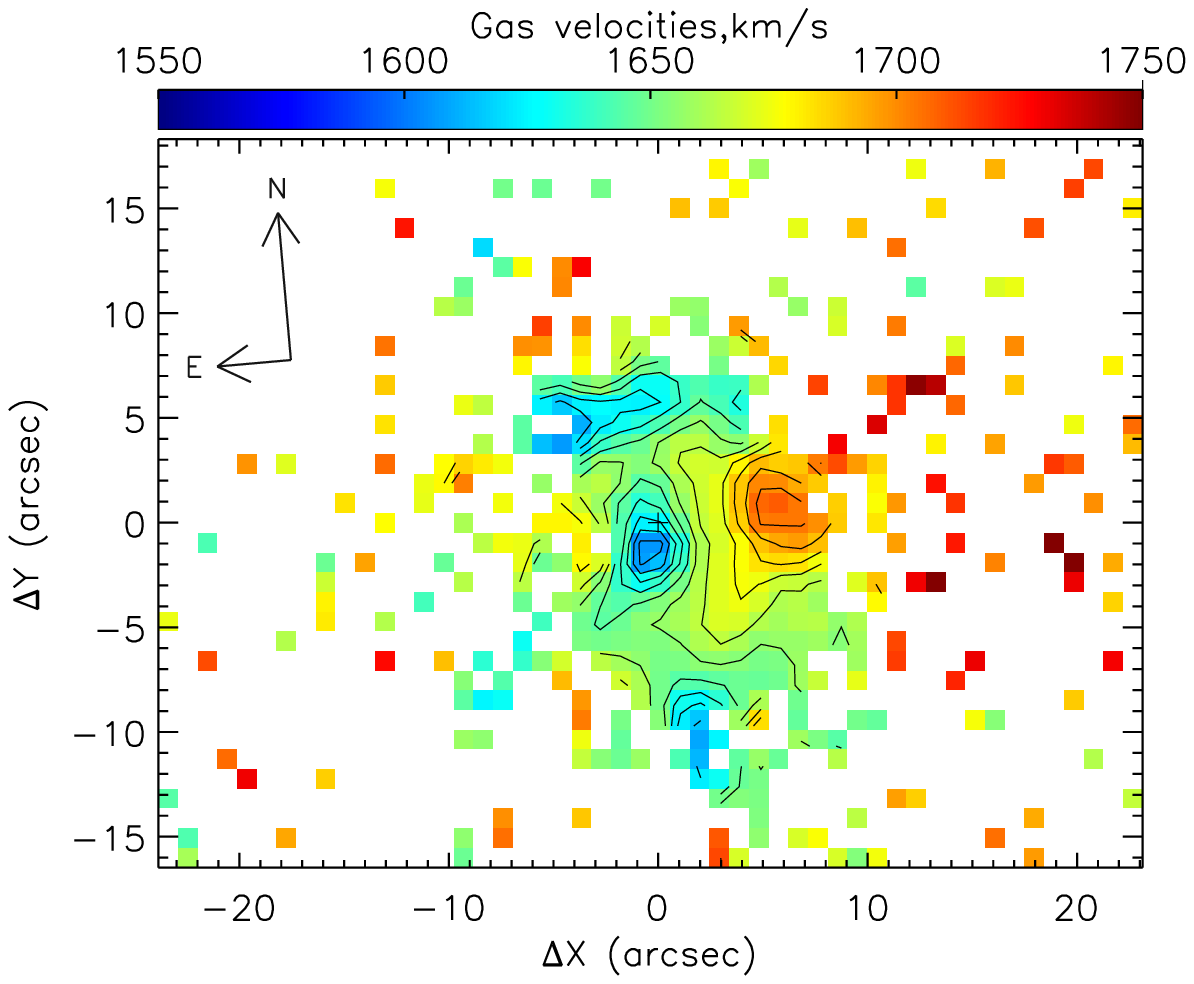}}
     \centerline{
       \includegraphics[width=0.33\textwidth]{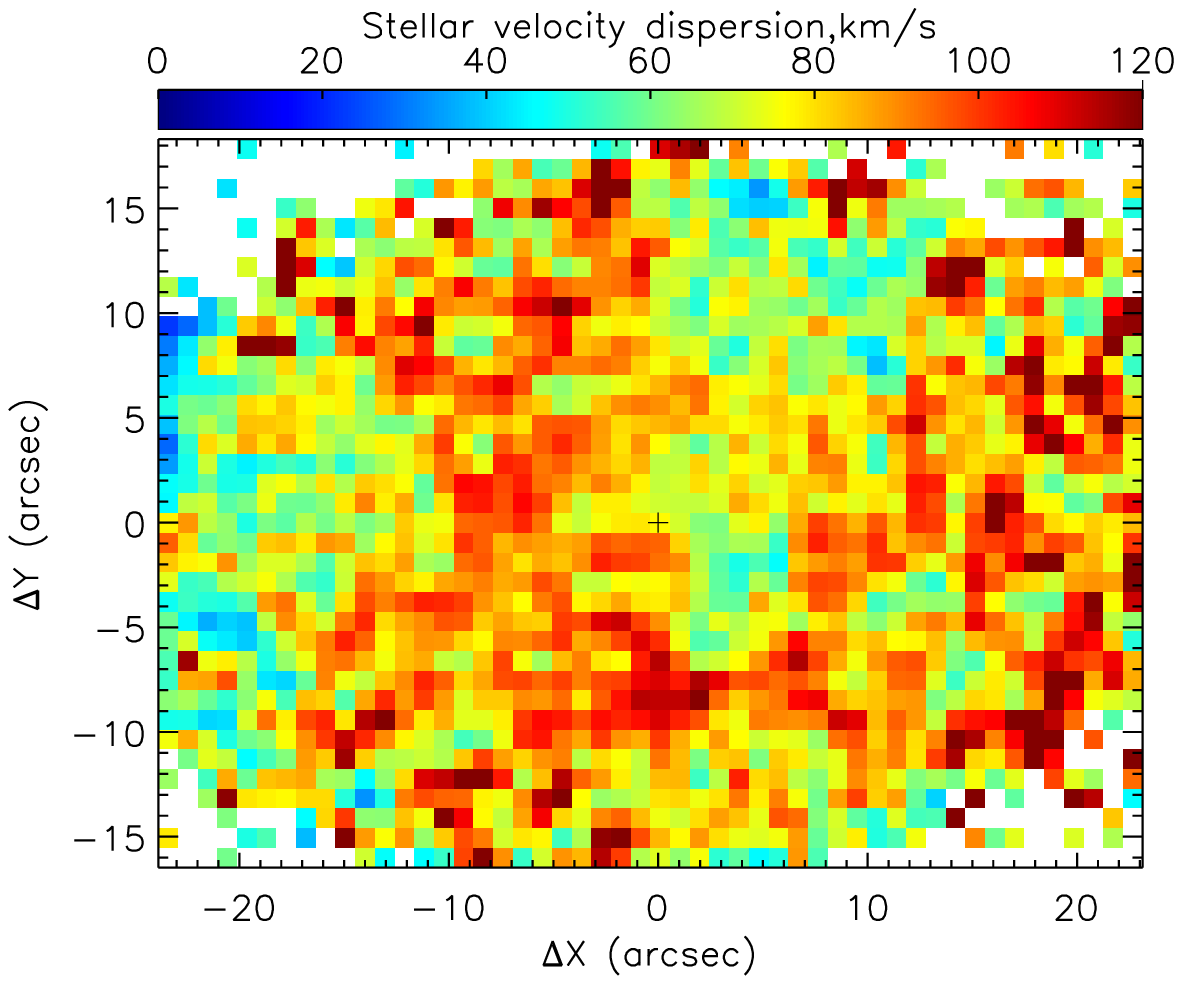}
       \includegraphics[width=0.33\textwidth]{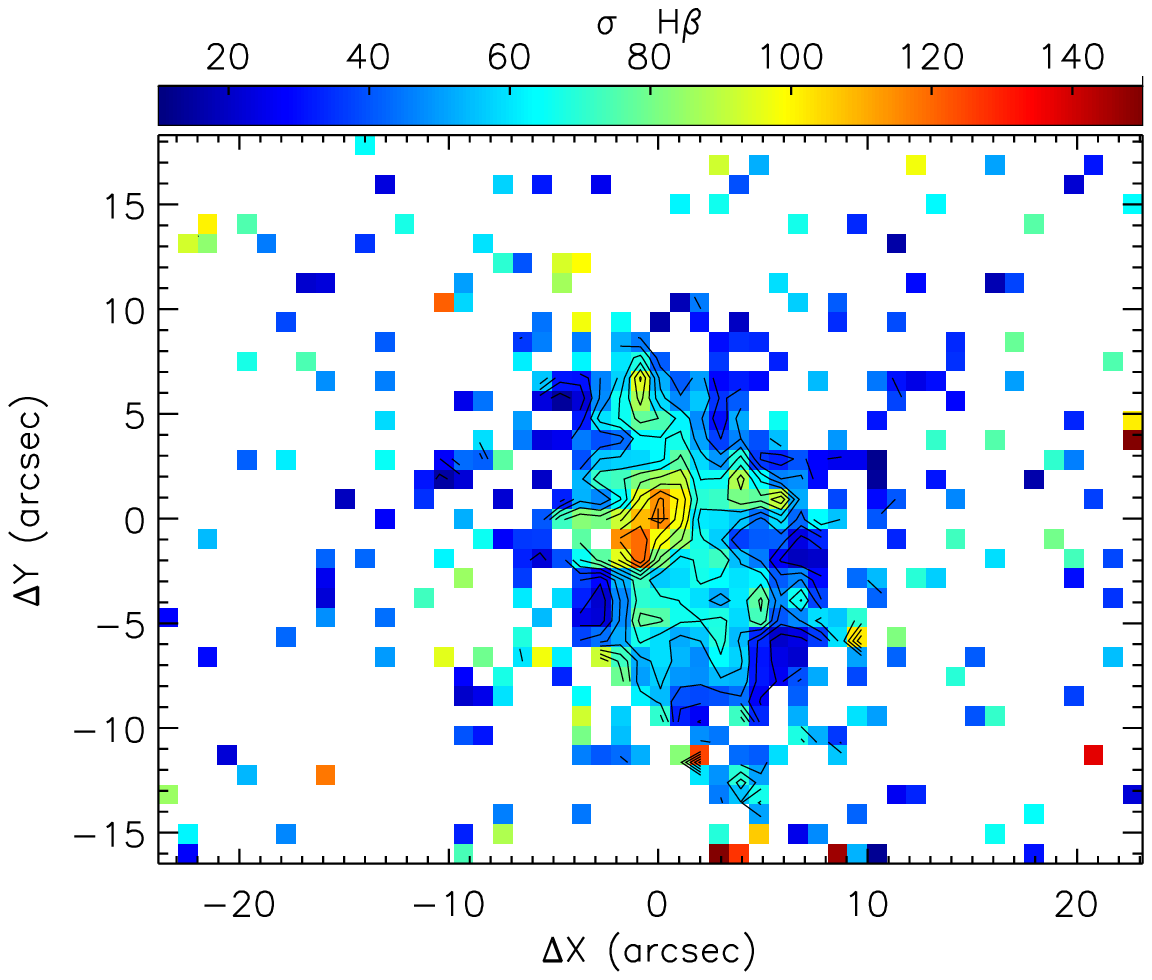}
       \includegraphics[width=0.33\textwidth]{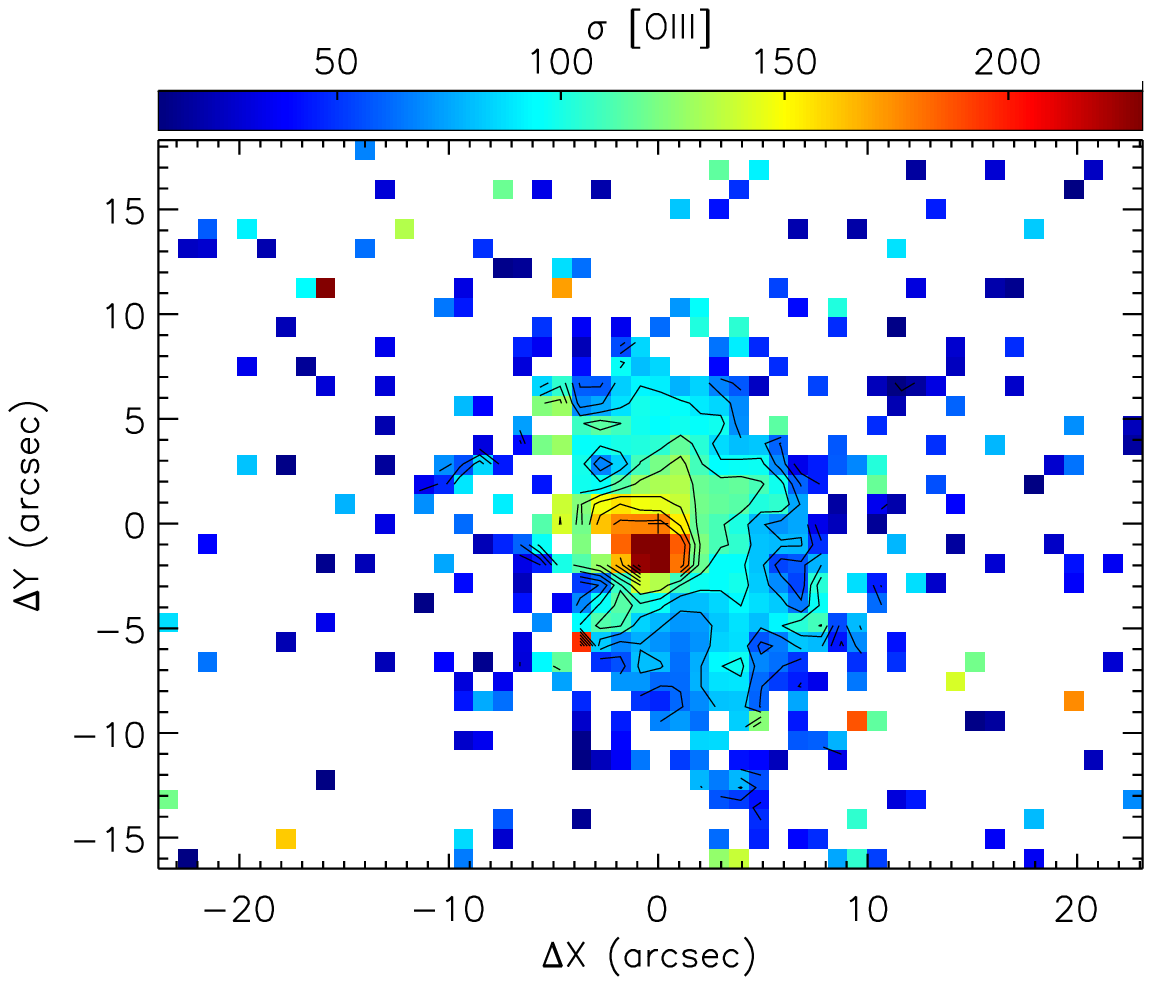}}
     \caption{The results of the SAURON data analysis. The top row shows the
 continuum image, the line-of-sight velocity fields for the  stars and for the
 ionized gas. The bottom row shows the map of the stellar component velocity
 dispersion and velocity dispersion maps for the lines  $H\beta$ and
 [OIII].}
     \label{fig_sauron_results}
 \end{figure*}

The data cubes were fitted with the ULySS software package in the same manner as
the long-slit data described above. The model of the LSF constructed from the
twilight sky spectra was taken into account. Because of the relatively low $S/N$
ratio of the emission lines we fitted simultaneously \Hb\, and [O\,III] emission
line profiles by assuming that their velocities are the same; the line widths
and intensities were free parameters. The maps were  smoothed by a Gaussian with
$FWHM = 1.5$ pixels. The resulting maps are presented in
Fig.~\ref{fig_sauron_results}.

\section{STELLAR CONTENT}
\label{sec_stars}

\subsection{Kinematics}
\label{stellar_kinematics}

Radial variations of the $v$ and  $\sigma$  derived from our long-slit data
(Fig.~\ref{fig_longslit_results}) show a good agreement with the SAURON maps
over the galaxy central region (Fig.~\ref{fig_sauron_results}). The
line-of-sight velocity curve at the
$PA=313\degr$ (which corresponds to the kinematical major axis, see below) has a
sharp small-amplitude internal maximum  at the  $R\approx 2\arcsec$ (190 pc). At
larger radii the rotation velocity  rises continuously up to
$R\approx 60\arcsec$ (5.5 kpc). The apparent amplitude of the stellar rotation
velocity is about  $160\km$.

\begin{figure}
\includegraphics[width=0.5\textwidth]{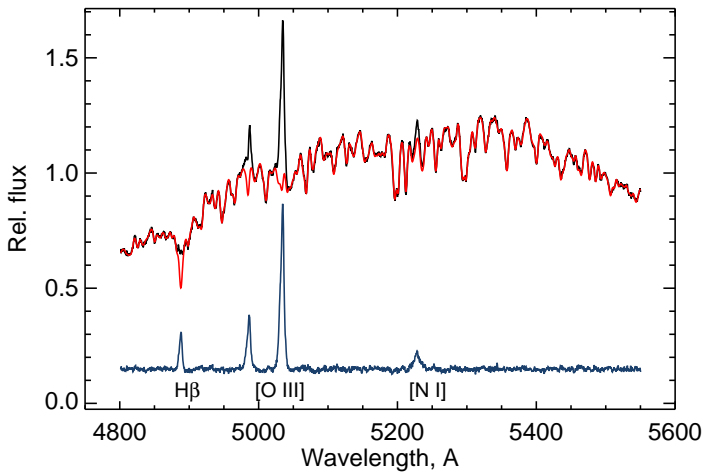}
\caption{A comparison between observed spectrum (black line) at $r=-2''$ into
cross-section $PA_{slit}=313\degr$ and model of stellar population (red line).
Blue line represents a residuals between observed spectrum and model one.}
\label{added_fig}
\end{figure}

We have assumed a purely Gaussian shape of the  LOSVD and ignored the
contribution of the $h_3$ and  $h_4$ moments, because the value of the intrinsic
velocity dispersion  is comparable to the width of the LSF ($\sigma_{LSF}\approx65\km$).
The $\sigma$ profiles along different $PAs$ are slightly different: the cross-sections
along the $PA = 313\degr$ and $358\degr$ reveal a peak of $80\km$ achieved in the
nucleus, while along the bar ($PA=268\degr$)  a central plateau has appeared.
A complex surface distribution of the $\sigma$ is confirmed by  the SAURON maps
(Fig.~\ref{fig_sauron_results}): in the central region  ($R<5$\arcsec) $\sigma$
is depressed with respect to the neighboring regions. Similar `sigma-drop' features
in barred galaxies  may result from a circumnuclear dynamically cold young stellar disks
\citep{Wozniak2003}.   \citet*{Koleva2008} have as well noted that  such
a $\sigma$-depression can be an artifact related to the template mismatch. However
in the case of NGC~7743 we propose interpretation of the $\sigma$-drop  in the
framework of the  \citet{Wozniak2003} concept, because  the presence of young
stellar population in the galaxy core  is confirmed  by our measurements (see
the next subsection).  We think also that in our case we have no problems with
template mismatch: the inspection of the residuals after the model spectra subtraction gives impression of the fitting quality (Fig.~\ref{added_fig}). Note that our estimate of the central velocity dispersion
is slightly lower than those of other authors: HyperLeda gives $84.5\pm 2.4\km$
averaged over six measurements.

The kinematical orientation parameters were determined by applying a model of
thin flat  circular rotation to the observed velocity fields.  The
model corresponds to $$ V_{obs}=V_{sys}+V_{rot}(R)\frac{\cos \varphi \sin i}{\sqrt{\sec^2 i - \cos^2 \varphi \tan^2 i}} $$
$$ R=r\sqrt{\sec^2 i - \cos^2 \varphi \tan^2 i}, $$
where $R$, $r$ -- is the radius in the galaxy plane and the radius in the plane of the sky, $i$ is the  inclination
of the galaxy plane. Position angle of the kinematical major axis of the stellar disk
(i.e. the disk line-of-nodes) $PA_0$ and angle $\varphi$ are related by $\varphi=PA_{slit}-PA_0$.
Rotation curve $V_{rot}(R)$ was determined with several fixed radial nodes and was then
interpolated to all radii using spline functions. Thus, the model parameters are the rotation
velocity values at  fixed radial nodes and the common inclination, and the line-of-nodes position angle.
 We  simultaneously fitted three long-slit  stellar velocity profiles (the most
extensive measurements being in the `green' domain) and the SAURON velocity
field. The best model has an inclination of $i_{star}=40\pm 2\degr$  and a major
axis position angle of $PA_{star}=310\pm 5\degr$.  The rotation curve is shown in
Fig.~\ref{compar_RC} (left column). \citet{Moiseev2004} have found
$PA_{star}=300$--$310\degr$ from the circumnuclear velocity field
of stars ($R<7$\arcsec),  which almost coincides with our new measurements for
the whole large-scale stellar disk.  It is important to note that the
large-scale bar could perturb  the circular velocities of stars, however we have
found that these perturbations are negligible in the case of  NGC~7743. First of
all,  it follows from the tilted-ring analysis of the SAURON velocity field
\citep[for the method description, see][]{Moiseev2004} that radial variations of
the kinematical major-axis $PA$ demonstrate very small deviations
($1$--$2\degr$) from the mean $PA_{star}$ (cited above) up to $R=25$\arcsec
(2.3 kpc).  Such a picture would have been impossible if the bar significantly
distorts the velocity field --  we should observe a turn of the $PA$ along the
radius because the bar potential contribution  decreases with the distance from
the center, while in the very center the rotation of the bulge dominates the
line-of-sight velocity field. On the other hand, we tried to search for the
orientation parameters  excluding the bar, area and  using only the elements
with $R> 30$\arcsec (2.8 kpc). This fit returns the value of $PA_{star}$
roughly equal to the estimates given above,  but the inclination cannot be
confidently determined in this fit.

\subsection{Age and metallicity}

Radial variations of the stellar population parameters  derived for NGC~7743
from our long-slit data -- those of the SSP-equivalent age and
metallicity -- are given in Fig.~\ref{fig_longslit_results}
together with the other radial profiles. The stellar nucleus of NGC~7743 has
appeared to be chemically and evolutionarily decoupled: the mean stellar age in
the nucleus is only 1~Gyr, the metallity is 1.5 times above solar. Beyond
the nucleus,   metallicity drops to subsolar values, and the age stabilizes at
about 2.5~Gyr. We suppose that our measurements at $R<20\arcsec -30\arcsec$
(i.e. 1.9--2.8 kpc) relate mainly to the bulge and bar of NGC~7743, based on
its surface brightness distribution \citep{Erwin2008}.

Earlier, we  already reported the chemical and evolutionary distinction of the
nucleus of NGC~7743 with our integral-field spectroscopic results for the
central part of the galaxy. By obtaining the data with the Multi-Pupil
Fiber Spectrograph (MPFS) of the Russian 6-m telescope,  \citet{Silchenko2006} has
measured the Lick indices and found that the nucleus was 5 times more
metal-rich, and  by more than 2~Gyr younger  with respect to the neighboring bulge.
However, the absolute values of the ages and metallicities obtained here
differ from those, obtained by  \citet{Silchenko2006}: there [Z/H]$=+0.7$ was
found for the nucleus and $0.0$ for the bulge, and the mean age
of the bulge was determined as 4~Gyr old. The emission lines were very
strong in the nuclear and circumnuclear spectra, and the excitation
mechanism of those was unclear, so perhaps the correction of the
Lick index H$\beta$ for the emission by \citet{Silchenko2006} was not completely
correct; hence, the stellar population parameters for NGC~7743 in Sil'chenko (2006)   may be
biased.
Later, \citet{SilchenkoChilingarian2011}  have analysed the SAURON data (which are
analysed here as well), by the method of direct spectrum fitting in the pixel
area, that allows to exclude all the spectral regions, affected by emission. The
results of the SAURON data analysis  by
\citet{SilchenkoChilingarian2011}
agree perfectly with the present results, derived by fitting the long-slit
SCORPIO spectra: they have obtained [Z/H]$=+0.2$ and the age of 1.2~Gyr  for
the nucleus, and [Z/H]$=-0.33$ and the age of 2.2~Gyr  for the bulge
\citep{SilchenkoChilingarian2011}
while Fig.~\ref{fig_longslit_results} implies [Z/H]$=+0.2$ and $T=1$~Gyr in the
nucleus and [Z/H]$=-0.3\pm 0.1$, and the age of 2.5~Gyr for the bulge.
Therefore, the data of several spectrographs give the evidence for the young age
and enhanced metallicity in the stellar nucleus of NGC~7743; we conclude that
there is a signature of a recent nuclear star formation burst.

In Fig.~\ref{fig_longslit_results} the stellar population parameter profile in
$PA=268\degr$, almost along the bar, is quite outstanding among the others. The
semiaxis of the bar is
$31\arcsec -37\arcsec$  (2.9--3.4 kpc), according to the estimates by
\citet{Erwin2005}.
At the slit orientation along the bar, unlike the other two profiles, the
metallicity does not
diminish toward the measurement limits, but stops its fall at the radius of
about
$30$\arcsec, near the end of the bar, and then even rises a little.
Perhaps, here we feel the influence of so called ansae -- the bright and often
starforming regions at the ends of a bar which are especially frequent in barred galaxies of
early  morphological types \citep{Laurikainen2007}: additional star
formation should increase simultaneously both a continuum surface brightness and
a mean stellar metallicity.

\begin{table*}
\footnotesize
\caption{Determination of the gaseous disk orientation.}
\begin{tabular}{ccccc|cc}
\tableline
\#   &   Emission lines      &  Cross-sections      &
$i$ (deg)	& PA (deg) & $\delta i_1$ & $\delta i_2$\\
\tableline
1	& all	& all	& 62	&	257	& 45$\pm$14 & 89$\pm$15	\\
2	& forbidden lines $[O\,III]$, $[N\,II]$, $[S\,II]$& all & 47 &
261 & 34$\pm$12 & 78$\pm$14\\
3	&  Balmer lines $H\beta$, $H\alpha$ & all & 42 &  273 &
24$\pm$11 & 77$\pm$15  \\
4	& all	&  $PA_{slit}=220, 268$  & 43 & 270 & 26$\pm$11 &
77$\pm$14\\
5 	& all   & $PA_{slit}=313, 358$ & 38 & 244 & 40$\pm$11 & 64$\pm$13\\
\tableline
\end{tabular}
\label{gas_model_parameters}
\end{table*}

\section{Ionized gas}
\label{sec_gas}
\subsection{Gas velocities}

Gas motions in NGC~7743 are more complex than those of the stars and look
decoupled in all position angles except in $PA=268\degr$ (
Fig.~\ref{fig_longslit_results}). Along the $PA=358\degr$
the velocity projections onto our line of sight have opposite winding senses for the gas and for the stars.
By using the SAURON data, we have constructed the well-sampled
2D ionized-gas velocity field, but only for the very inner  region,
$R<8$\arcsec\  (740 pc). In the meanwhile, due to the long exposures and to
adaptive binning
application, with the SCORPIO/long-slit we have derived emission-line velocities
up to the edge of the stellar velocity measurements -- toward $60$\arcsec\
(5.6 kpc) from the center.

Close to the minor axis of the stellar disk ($PA=220\degr$), where the stellar
rotation velocity projections are almost zero, the gas line-of-sight velocity
oscillates with the amplitude of $50-70\km$. Inspecting the SAURON--based gas
velocity field, we see a `blue' velocity spot in $1\arcsec -2\arcsec$
(i.e. 90--190 pc) to the south from the nucleus; this peculiar region is to be
discussed below, in the Subsection 5.3. Trying to reconcile the gas velocity profiles
in four position angles, we come to the conclusion that the ionized-gas motions inside
the central kiloparsec are quite different from those in the outer disk.
Within $R<10\arcsec$  (900 pc) at least two mechanisms affecting the
gas motions can be identified: the first, an active nucleus impact, and the
second, non-circular velocities induced by the circumnuclear spiral wave which settles
perhaps off the stellar disk plane. At the same time, the outer gas rotates rather
regularly, though in other plane than the stars do.

We have performed a series of calculations of  the circularly rotating model of a thin disk for the outer gas (at $R=25\arcsec -50\arcsec$,  2.3--4.7 kpc)  alternately excluding different
cross-sections and emission lines, using the same formulae as described
in the Section~\ref{stellar_kinematics}.

\begin{figure*}
\includegraphics[width=\textwidth]{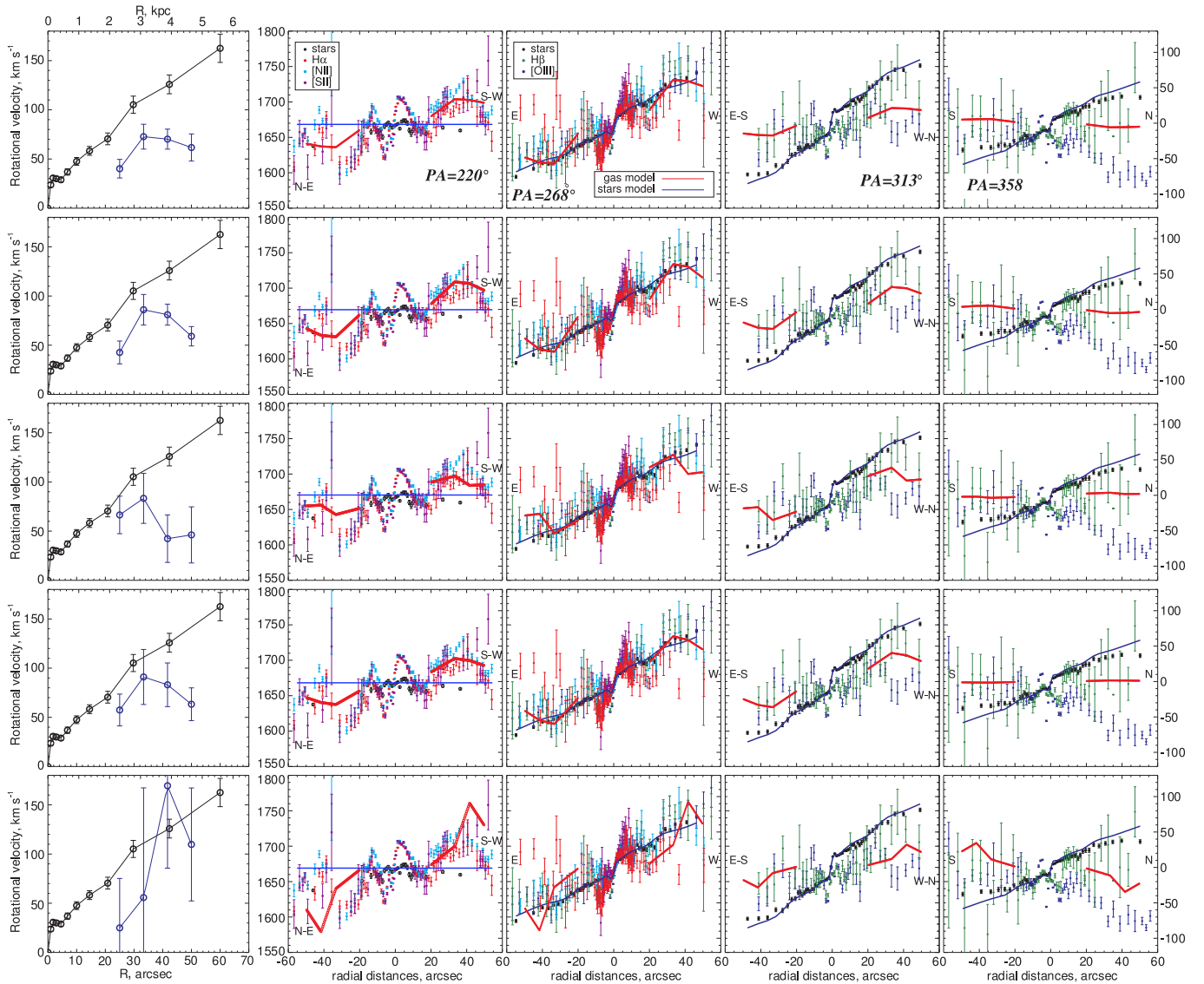}
\caption{Series of experiments to determine the circularly rotating
model of a thin disk to the outer gas. Each raw corresponds to the
model configuration  described in the Table~\ref{gas_model_parameters} -- \#1 to 5 from top to bottom, respectively. The left
column shown the stellar rotational model (black lines and circles) and the
gaseous one (blue lines and circles). The following columns correspond to
different cross-sections. Black circles present line-of-sight stellar velocity,
color circles -- gaseous velocity. The blue and red lines show the
stellar and gaseous rotation model projections to  different cross-sections,
respectively. }
\label{compar_RC}
\end{figure*}

The  considered configurations (combinations of the emission lines in different cross-sections) are listed in the Table~\ref{gas_model_parameters}. Fig.~\ref{compar_RC} shows the comparison of the model rotation curves (blue lines) with the observed line-of-sight velocities for every long-slit cross-section. The first configuration (the top row of the Fig.~\ref{compar_RC}) involves all available cross-sections. It shows a good agreement with the data in $PA=220\degr$   and $PA=268\degr$, however it has a mismatch with the observed velocities in W-N ($PA=313\degr$) and N ($PA=358\degr$)  sides of the galaxy disk. The main problem is significant differences of the velocities estimated in the $H\beta$ and $[OIII]$ emission lines (see discussion below). From this reason we have also considered  separately fitting for the forbidden ([OIII], [SII], [NII])  and Balmer ($H\alpha$, $H\beta$) emission lines, see the corresponding configurations \#2  and \#3. Note that resulting rotation curve in the case \#2 seems to be more realistic comparing with an abrupt decrease of the rotation velocities for \#3. It may be related to the fact that some part of the Balmer emission is produced by the main disk (see below). Also we have considered the numerical stability of our fitting using different combinations of cross-sections. In the configuration \#4 we excluded points  along $PA=313,358\degr$ with the most dramatic difference between the ionized gas kinematics in the forbidden and Balmer lines. On the contrary, in the configuration \#5 only points from the $PA=313,358\degr$ cross-sections were used.

Further, we determine the angle
between the stellar and gaseous disks  using the following
formula  \citep{Moiseev2008}:
$$
\cos \delta i = \cos (PA_{gas} - PA_{star}) \sin i_{star} \sin i_{gas} +
\cos
i_{star} \cos i_{gas}.
$$
Here we assume that $i<90\degr$ if the angular momentum vector is directed towards the
observer, while in the opposite momentum direction the inclination is $180\degr - i$. Supposing that the large-scale spiral arms in the stellar disk are trailing, we determine $i_{star} = 40\degr$.
However, for the gaseous disk we cannot choose between the alternate directions
of its angular momentum vector, hence  for the mutual inclination angle we obtain
two possible solutions, $\delta i_1$ and  $\delta i_2$. Table~\ref{gas_model_parameters}
presents the resulting parameters of different configurations.

 The mutual inclination angles for all configurations considered here are in agreement with each other within the
error bar. In any case (see the two last right columns of Table~2), the large-scale gaseous disk
appears to be strongly inclined to the main symmetry plane of the galaxy. The mean values for these angles, averaged over all our experiments, are  $\delta i_1=34\pm9\degr$, or $\delta i_2=77\pm9\degr$. Here the errors correspond to rms.  We have adopted these values as the possible mutual inclination angle between the stellar and the gaseous disks.

Meanwhile, the difference between the rotation planes of the gas, emitting
mostly in the forbidden lines (shock-excited gas) and that, emitting prominently
in the Balmer lines (star-formation excited gas) may be real. Also, we think that
the difference in the orientation parameters for the two groups of the cross-sections
(the experiments number 4 and 5 in the Table~2) may be physically motivated: perhaps,
to the south from the line-of-nodes  at $PA_{gas}=220\degr -270\degr$, we mainly see
the inclined, shock-excited gaseous disk, and to the
north from the line-of-nodes, this inclined gaseous disk is seen through the own
gas-poor disk of NGC~7743, possible illuminated  by a small amount of young stars.

 \begin{figure*}
 \centerline{
 \includegraphics[width=0.5\textwidth]{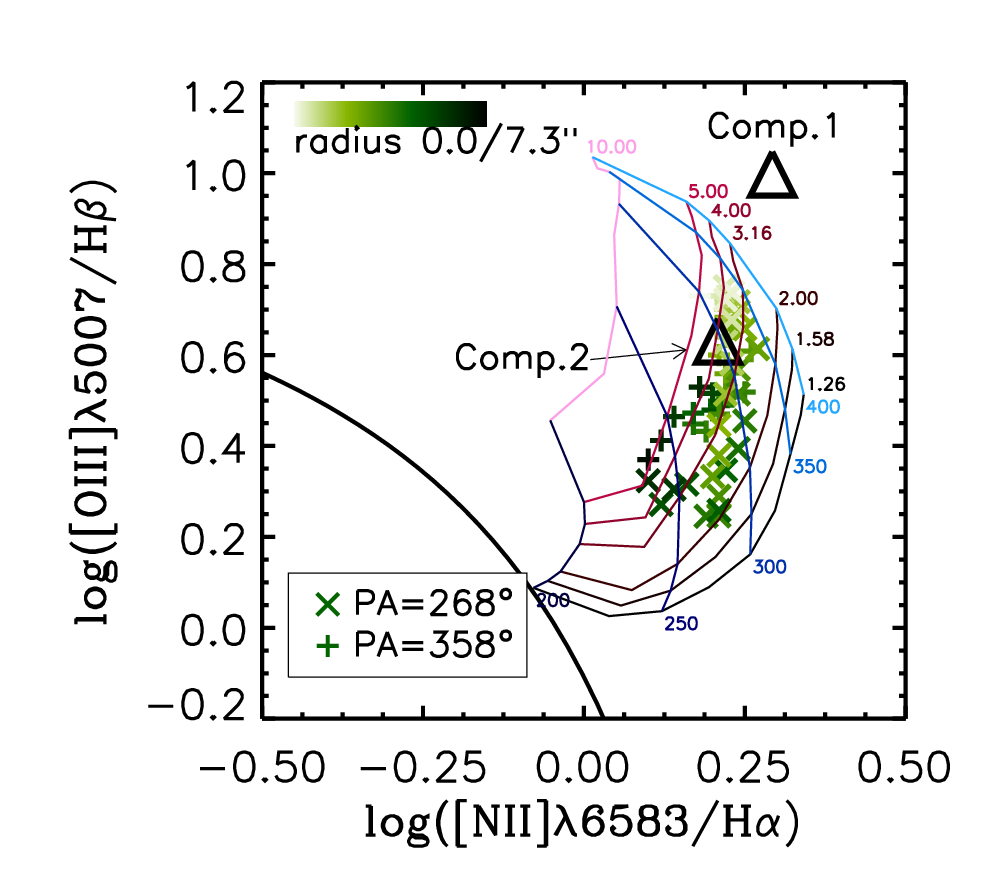}
 \includegraphics[width=0.5\textwidth]{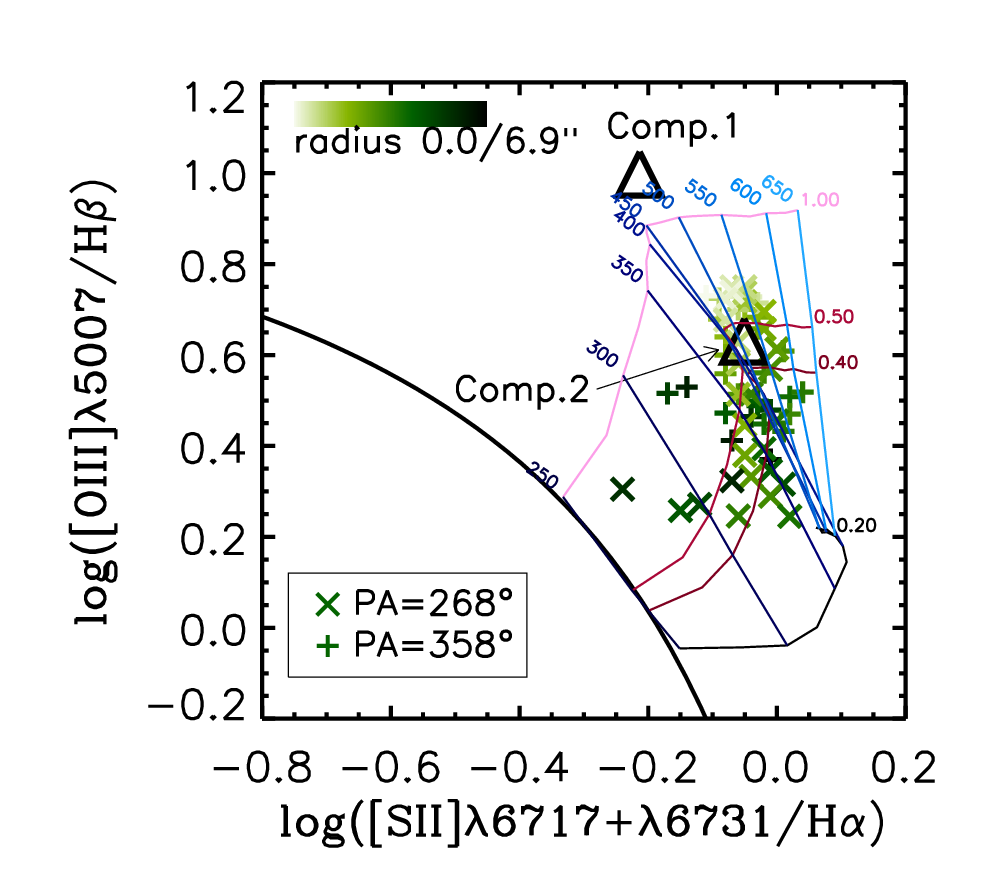}
 }
 \caption{The excitation diagnostic diagrams confronting the emission-line
intensity  ratios: $\mbox{[O\,III]} / \mbox{H}\beta$ vs $\mbox{[N\,II]} / \mbox{H}\alpha$
({\it left})  and $\mbox{[O\,III]} / \mbox{H}\beta$ vs $\mbox{[S\,II]} / \mbox{H}\alpha$
({\it right}).  The pluses and crosses mark different long-slit cross-sections. The intensity of
the green color corresponds to the increasing distance from the center in the sky plane. The
solid black  curve, which separates the areas with the AGN-induced and star formation
induced excitations,  is taken from \citet{Kewley2001}. Our measurements are overlaid by the
shock-excitation  model \citep{Allen2008} for the electron density of $n_e = 0.1\cm$ and
solar element  abundances. The red colour traces magnetic parameter isolines,   blue   --
the shock  velocity isolines. Black triangles mark two separate emission-line components
for the regions  with   clear emission-line splitting; `comp.1' relates to the blueshifted
component,  `comp.2' -- to the main one.
 }
 \label{diagn_diagramm}
 \end{figure*}

\subsection{Gas ionization sources}

In order to identify the source of gas ionization, we plot our measurements for
the inner part of the galaxy, where several emission lines of various excitation
degrees are measurable, onto the classical excitation-type diagnostic diagrams
(Fig.~\ref{diagn_diagramm}). We do not need to take into account the intrinsic
reddening as we use the intensity ratios of close pairs of emission lines.
One can see (Fig.~\ref{diagn_diagramm}) that all the measurements are located
in the AGN-ionization area of the diagrams. With the increasing distance from
the center, the contribution of the non-thermal source into the gas excitation
diminishes
 because the [OIII]$\lambda$5007 intensity decreases and the observed points
cross the border between the AGN-type and LINER-type excitation
\citep[$\mbox{[OIII]} / \mbox{H}\beta =3$][]{VeilleuxOsterbrock1987}.
The off-center measurements can be fitted best of all by recent models of shock
excitation for the low-density gas \citep[{\it shock$+$precursor},][]{Allen2008}
if one supposes a moderate value of the magnetic parameter $B$ and the speed of
the shock wave of more than 250$\km$.

Therefore, while in the  vicinity of the nucleus the gas ionization is mainly produced by an
active non-thermal radiation source, just in 300--500 pc from the center the main
contributors into the ionization are the shock waves. Perhaps, one of the local producer
of the shock wave is the jet of the active nucleus (discussed below). For the
more distant parts of the gaseous disk it is reasonable to relate   shock waves
to the inclined orbits of the gas clouds, which must collide with the interstellar
medium of the main large-scale stellar disk (even if the density of the latter is very low).
Moreover, a shock wave may be induced by a simple gas cloud, crossing   the
potential well of the stellar disk, quite analogous to the shock being induced by a gas cloud,
crossing  a spiral density wave in large-scale disks of spiral galaxies.
 This mechanism was discussed by \citet{Wakamatsu1993}, who considered the shock wave
generation in the gas of polar ring galaxies.  In favor of the latter hypothesis we can
refer to the measurements along $PA=220\degr$, where we trace the emission-line
intensity ratios towards the largest distance from the center. At the radii of
$R=20\arcsec -55\arcsec$  (1.9--5.1 kpc)  we obtain
$\log \mbox{[NII]} / \mbox{H}\alpha >-0.3$ that excludes  photoionization by
young stars \citep{Stasinska2006}. Therefore, throughout the whole large-scale gaseous disk
of NGC~7743 the gas is ionized by shock waves.

\subsection{Jet-clouds interaction}
\label{agn_jet}

\begin{figure*}
 \includegraphics[width=\textwidth]{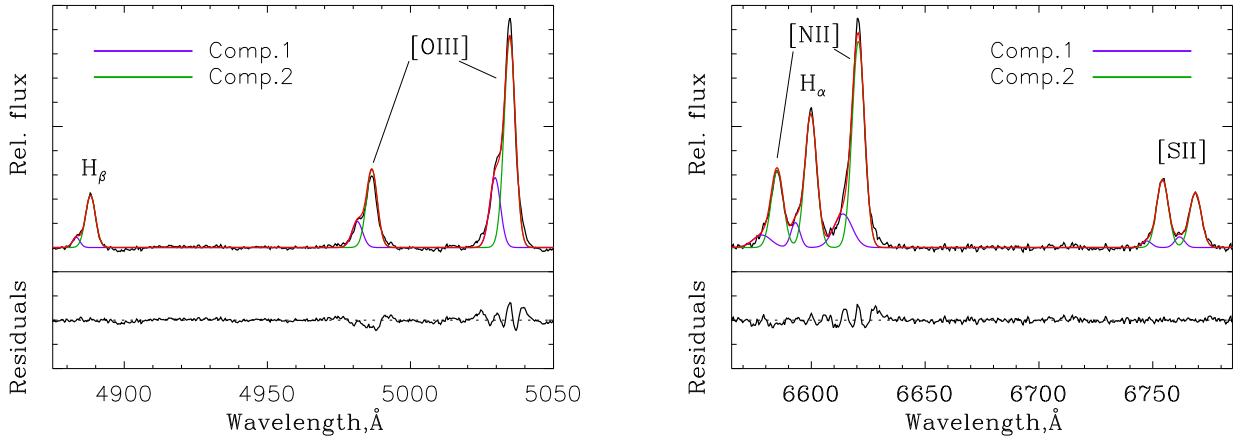}
 \caption{The multi-component emission-line profile, obtained by integrating the
long-slit  data within the range of $R=-4$--$0$\arcsec along the cross-section in
$PA=358\degr$.  The black line shows the observed emission-line spectrum, after subtracting
the model  of stellar spectra. The blue line corresponds to the Gaussian-profile model of
 the blueshifted component (`Comp. 1'), the green line -- to the
Gaussian-profile model  of the main emission component (`Comp. 2'), and the red line traces the full
model  profile.}
 \label{multicomponent_profile}
 \end{figure*}

Fig.~\ref{fig_longslit_results} shows that the velocity dispersion of ionized
gas in the center of NGC~7743 exceeds that of the stellar component  more than
twice; evidently, it is an active nucleus effect. Moreover, in the long-slit
cross-sections along $PA=313\degr$ and $PA=358\degr$ the gas velocity dispersion
peaks are displaced from the (optical) nucleus by 1\arcsec --2\arcsec\   (i.e. by 90--190 pc),
and the gas line-of-sight velocity drops at this radius. The SAURON maps reveal this
displacement even better: in Fig.~\ref{fig_sauron_results} one can see a `blue'
velocity spot and the gas velocity dispersion peak at $2$\arcsec\ to the
south from the nucleus. Within this `spot' the emission-line profiles in the
SAURON spectra look asymmetric, and the SCORPIO spectra, which have  better spectral
resolution allow to decompose the entire  profiles into two emission-line
components (Fig.~\ref{multicomponent_profile}). We  simultaneously fitted the emission-line
profiles for $\mbox{H}\beta$, [O\,III], [N\,II], \Ha , and [S\,II] via a set of
Gaussians broadened with the LSF, by requiring the same velocity difference
between two components for all species. The best fit is presented in
Fig.~\ref{multicomponent_profile}; the more intense component of all the lines
(Comp.~2) corresponds to the regularly rotating gas, and the faint emission-line
components (Comp.~1) are blueshifted by $299 \pm 8 \km$. Big black triangles in
Fig.~\ref{diagn_diagramm} show intensity ratios for two components separately. One can see that the ratio
of $\mbox{[OIII]} / \mbox{H}\beta $ for the off-nuclear Comp.~1 is higher than this
ratio in the whole disk, and it is even higher than that in the close vicinity of the
active nucleus. A similar picture -- a high-speed strong shock excitation
along  with  non-circular motions of the gas in a local off-centered area -- has been observed by
\citet{Smirnova2010a}  in the Seyfert galaxy Mrk~334, where this area has been identified as
a point of a satellite fragment crossing the galaxy disk. Another consequence of
such crossing must be the lower gas density in this `hole' with respect to the
neighboring disk. However in NGC~7743 on the contrary,  the blueshifted Comp.~1  reveals a
very high electron density, $3000 \pm 1500\cm$, which is
determined from the intensity ratio of $\mbox{[S\,II]}\lambda 6717 /
\mbox{[S\,II]}\lambda 6731$, while the regularly rotating disk Comp.~2 only has $n_e=150 \pm 30 \cm$.
Therefore, in this peculiar region we detect gas compression at the shock wave front. The
probable cause of the shock waves can be an active-nucleus jet, penetrating the
surrounding interstellar medium. Such intrusions are observed in many Seyfert galaxies,
for example, in Mrk~3 \citep{Capetti1999}. Unfortunately, for NGC~7743 we have no narrow-filter
emission-line images with a sufficiently good spatial resolution to inspect this
peculiar region in detail. But \citet{ReganMulchaey1999} mentioned a `blue-color
excess' just in $1$\arcsec  ( $\sim90$ pc) to the SE from the nucleus at the HST
map of the galaxy -- it may be the region of the suggested jet intrusion. In this case, we only see the
jet directed towards us, while the counterjet is obscured by the galaxy disk.

The radio image of the jet would be a decisive proof for this suggestion. But, as
we noted in Section~2, all the radiocontinuum maps of NGC~7743 demonstrate only
its compact core. Perhaps, the spatial resolutions and sensitivities of these
radio observations are insufficient to resolve a jet.

\section{Discussion}
\label{sec_dis}

The previously known data (see Section~\ref{sec_n7743}) characterized NGC~7743
as an ordinary early-type spiral galaxy with a weak nuclear activity. The traces of
nuclear jet intrusion (the size of the jet must be less than 200 pc) into the
interstellar medium, which we have found, are not so surprising for a Seyfert
galaxy. However, another astonishing fact is revealed: the bulk of the ionized gas, up
to the distance of 4--5 kpc from the center, rotates on the orbits strongly
inclined (by $\delta i =34\degr$ or by $78\degr$) to the plane of the stellar disk.

The large-scale disk of NGC~7743 lacks prominent HII-regions, and the
gas excitation determined by us from the emission-line spectrum is  shock-like.
A similar situation has been observed by us in NGC~5631 \citep{Silchenko2009},
 where the large-scale gaseous disk is slightly inclined to the stellar
one, and has an opposite spin. We have suggested two possible
origins of the shock-like excitation: it can be either due to the collisions between the
accreted gaseous clouds on the inclined orbits with the original low-density gaseous disk
coplanar to the stellar one, or due to the accreted gas clouds, crossing the gravitational
well of the stellar disk. However, in any case the gaseous content of the main
galactic disk must be several times smaller than the accreted gas mass. Otherwise
we would see two-component emission-line profiles over the whole disk extension,
not only in the jet region.

Unfortunately,  we have not found any simulations of shock
waves induced in inclined gaseous disks in the literature,  except
the  estimations by \citet{Wakamatsu1993} for the  case of polar rings. However, qualitatively we consider the mechanisms,
proposed above as quite probable and similar to the spiral density wave
generation in large-scale gaseous disks of spiral galaxies and to the
shock-induced dust lanes at the leading edges of bars. These scenario can also
explain the complex gas velocity behavior in the inner part of NGC~7743: the large-scale
bar of the galaxy does not affect the gas at large distances from the center,
but can affect its rotation at $R<20$\arcsec.

Among the galaxies known as `polar-ring galaxies' we can note several cases
of strongly inclined rings. First of all, it is NGC~660 where the outer gas-dust
ring is inclined by $\delta i =63\degr$ to the main galactic disk
\citep{ArnaboldiGalletta1993}. An inclined gaseous disk, less pronounced than that in NGC~660, and more similar
to the case of NGC~7743  has been recently found in Arp~212 by \citet{Moiseev2008}.
In Arp~212, a set of HII-regions within the radii of 2--6 kpc concentrates in  the
plane, which is inclined by $\delta i =30\degr$--$50\degr$ with respect to the inner
disk of the galaxy; in the regions where two disks are crossing,  prominent shock fronts are
seen. \citet{Moiseev2008} suggested gas accretion from the neighboring gas-rich
satellite, as the most probable mechanism of the inclined disk formation in Arp~212.
Such a scenario for the formation of polar and inclined gaseous rings was firstly
proposed by \citet{ReshetnikovSotnikova1997}  and greatly developed during the past
years, see \citet{BournaudCombes2003}. We think that a similar scenario has provoked
formation of the gaseous disk in NGC~7743. An alternate explanation may be minor merging
or even a disruption of a gas-rich satellite by  tidal forces from NGC~7743. But the
latter scenario would produce other consequences, such as kinematically decoupled stellar
subsystems, or the outer shells and arcs, which are not observed in NGC~7743. We hence conclude
that in any case a possible contribution of the accreted stellar component in NGC~7743
would be small, and we have to come back to the gas-dominated accretion as the probable cause
of the present state of NGC~7743.

In Section~\ref{sec_n7743} we have already mentioned that near NGC~7743 there
are at least two gas-rich satellites, so there is no problem to identify a source of
gas accretion. Moreover, in $50^{\prime}$ (279 kpc) from NGC~7743, another
S0-galaxy, NGC~7742, is located; the systemic velocity of NGC~7742 differs from that of
NGC~7743 by only $50\km$. Let us imagine that earlier NGC~7742 and NGC~7743 approached
each other with a typical relative velocity of $200\km$; then the time spent after
this passage is about 1.4~Gyr. This time is comparable to the age of the nuclear star
formation burst in NGC~7743, that may be induced by this interaction. However, we
doubt that the inclined gaseous disk was formed during the same event, because
this inclined-disk configuration is dynamically unstable and cannot persist its
inclined rotation during several periods. It is probable that the inclined-orbit gas
was accreted more recently.

\section{Conclusions}

Using the data of deep long-slit spectroscopy, obtained at the Russian 6-m
telescope, and  the archive data of the integral-field spectrograph SAURON, we have
studied the stellar population properties and kinematics of the stellar and ionized-gas
components in the early-type disk galaxy NGC~7743. We report the following
results.

\begin{itemize}

\item{All the ionized gas at radii of 1.5--5.4 kpc is confined to the disk,
inclined strongly to the main stellar disk of NGC~7743. The angle between two disks is
estimated; two possible solutions are obtained, $34\pm9\degr$ or $77\pm9\degr$,
depending on the mutual disk orientation in the space. The most probable origin of this inclined
gaseous disk is the accretion from the gas-rich environment of NGC~7743. The main
contributor into the gas excitation are the shock waves, induced probably by inclined-orbit gas clouds,
crossing the main stellar disk.
}
\item{Complex motions, multi-component emission-line profiles, and strong shock
excitation of the gas inside a compact region in 1\arcsec --2\arcsec\
to the south from the nucleus are interpreted by us as the signatures of
active-nucleus jet interaction with the surrounding ambient medium.
}
\item{We confirm the recent, about 1~Gyr ago, intense star formation burst in
the very center of NGC~7743, perhaps stimulated by the tidal interaction with the
neighboring NGC~7742.
}
\end{itemize}

The inclined gaseous disk in NGC~7743 has one of the lowest contrasts among the
structures of this kind, discovered in external galaxies. We think that deep
spectroscopy of other gas-poor early-type disk galaxies would allow
to find a lot of such inclined gaseous disks, formed by gas accretion from
external sources.

\acknowledgments

This research is partly based on the data, obtained from the Isaac Newton
Group Archive which is maintained as part of the CASU Astronomical
Data Centre at the Institute of Astronomy, Cambridge. We have made use of the NASA/IPAC Extragalactic Database (NED),
which is operated by the Jet Propulsion Laboratory, California Institute of Technology, under
contract with the National Aeronautics and Space Administration. We acknowledge
the usage of the HyperLeda database. This work was supported by the Russian
Foundation for Basic Research (project no.~09-02-00870). AVM is also grateful to
 the `Dynasty' Fund.  The authors thank the anonymous referee for constructive advice that has helped us to
improve the paper.

\end{document}